\documentclass[prl,twocolumn,superscriptaddress,floatfix]{revtex4}

\usepackage{amsmath}
\usepackage{amssymb}
\usepackage{graphicx}
\usepackage{dcolumn}
\usepackage{times}

\begin{document}

\title{Topological change of the Fermi surface in ternary iron-pnictides \\
with reduced $c/a$ ratio:
A dHvA study of CaFe$_2$P$_2$}

\author{Amalia I. Coldea}
\affiliation{H.H. Wills Physics Laboratory, University of Bristol, Tyndall Avenue, Bristol, BS8 1TL, UK}

\author{C.M.J. Andrew}
\affiliation{H.H. Wills Physics Laboratory, University of Bristol, Tyndall Avenue, Bristol, BS8 1TL, UK}

\author{J.G. Analytis}
\affiliation{Stanford Institute for Materials and Energy Sciences, SLAC National Accelerator Laboratory, 2575 Sand Hill
Road, Menlo Park, CA 94025, USA} \affiliation{Geballe Laboratory for Advanced Materials and Department of Applied
Physics, Stanford University, USA}

\author{R.D. McDonald}
\affiliation{Los Alamos National Laboratory, Los Alamos, NM 87545, USA}

\author{A.F. Bangura}
\affiliation{H.H. Wills Physics Laboratory, University of Bristol, Tyndall Avenue, Bristol, BS8 1TL, UK}

\author{J.-H. Chu}
\affiliation{Stanford Institute for Materials and Energy Sciences, SLAC National Accelerator Laboratory, 2575 Sand Hill
Road, Menlo Park, CA 94025, USA} \affiliation{Geballe Laboratory for Advanced Materials and Department of Applied
Physics, Stanford University, USA}

\author{I.R. Fisher}
\affiliation{Stanford Institute for Materials and Energy Sciences, SLAC National Accelerator Laboratory, 2575 Sand Hill
Road, Menlo Park, CA 94025, USA} \affiliation{Geballe Laboratory for Advanced Materials and Department of Applied
Physics, Stanford University, USA}

\author{A. Carrington}
\affiliation{H.H. Wills Physics Laboratory, University of Bristol, Tyndall Avenue, Bristol, BS8 1TL, UK}

\begin{abstract}
We report a de Haas-van Alphen effect study of the Fermi surface of CaFe$_2$P$_2$ using low temperature torque magnetometry up to 45~T.
This system is a close structural analogue of the
collapsed tetragonal non-magnetic phase of CaFe$_2$As$_2$.
We find the Fermi surface of CaFe$_2$P$_2$ to differ from other related
ternary phosphides in that its topology is highly dispersive in the {\it c}-axis,
being three-dimensional in
character and with identical mass enhancement on both electron and hole pockets ($\sim 1.5$).
The dramatic change in topology of the Fermi surface suggests that in a
state with reduced ($c/a$) ratio, when bonding between pnictogen layers becomes important,
the Fermi surface sheets are unlikely to be nested.

\end{abstract}


\date{\today}
\maketitle

The nature of the Fermi surface dimensionality and its proximity to
nesting in the iron-pnictide superconductors is at the core of
understanding the mechanism of superconductivity. An important
finding is that under {\it hydrostatic} pressure, BaFe$_2$As$_2$ and
SrFe$_2$As$_2$ \cite{Alireza08} superconduct whereas CaFe$_2$As$_2$
does not \cite{YuAWBNCL09}. Instead it undergoes a structural
transition to a phase with a much reduced $c$-axis length which is
known as the `collapsed tetragonal' (cT) state. If pressure is
applied using a non-hydrostatic medium CaFe$_2$As$_2$ does become
superconducting \cite{TorikachviliBNC08,Alireza08}.  This suggests
that small uniaxial pressure components stabilize the
superconductivity in CaFe$_2$As$_2$. Understanding why this happens in these
materials and whether
the Fermi surface nesting or the strong coupling with the lattice are the relevant
parameters will be important for understanding the
origin of superconductivity in iron pnictides.

CaFe$_2$As$_2$ has three distinct magnetic
and structural phases. Like many other iron-arsenides, at zero pressure, the high temperature
tetragonal state becomes orthorhombic and antiferromagnetically ordered below $T_N\simeq 170$\,K.
At low temperatures under a
pressure of 0.35\,GPa there is a transition to the cT state
in which there is a $\sim 10$\% decrease in the
$c$-axis lattice parameter and a $\sim 2$\% increase in the $a$-axis  \cite{KreyssigGLSZLBTNNLPAHMCG08}.
 Recent bandstructure calculations
\cite{Zhang_BS08,Tompsett09} suggest that these phase transitions have a dramatic effect on the Fermi surface. In the high
temperature tetragonal phase the Fermi surface is predicted to be similar to the other 122 pnictide consisting of two
strongly warped electron cylinders at the Brillouin zone corners and hole cylinders at the center of the zone
($\Gamma$). Calculations suggest that these hole pockets are sensitive to structural details of the As position and the
$c$-lattice constant, whereas the electron pockets are less affected \cite{Zhang_BS08}.
However, in the cT state the calculated Fermi surface suffers a major topological change;
the two electron cylinders at the zone corners
transform into a single warped cylinder whereas the hole cylinders
become a large three dimensional
sheet.

The Fermi surface of many iron-pnictides have been extensively studied by angle resolved photoemission spectroscopy,
however so-far the cT phase of CaFe$_2$As$_2$ has been inaccessible to this technique because of the high pressures
involved.  Quantum oscillation (QO) studies have the advantage of probing precisely the bulk three dimensional Fermi
surface but so far have only been possible in the magnetically ordered phase of the iron-arsenides \cite{Analytis08} as the
tetragonal superconducting phase cannot be accessed because of the high
values of the upper critical field ($\sim 100$~T).  Although QO studies are
possible under high pressure, these measurements are technically very challenging. One route to provide an insight into
the Fermi surface properties of the iron-arsenides is to study the analogous non-magnetic phosphide materials.
These materials have almost identical calculated Fermi surfaces to the arsenides in their non-magnetic state but are either
non-superconducting or have a low $T_c$ (and low $H_{c2}$) and single crystals
can be grown with very high purity making them ideal for QO studies. Quantum oscillations
have been measured in superconducting LaFePO \cite{Coldea08}
(which is analogous to the non-magnetic `1111' arsenides e.g., LaFeAsO) and SrFe$_2$P$_2$
\cite{Analytis09} (which is a structural analogue to the non-magnetic tetragonal `122' arsenides).

\begin{figure}[htbp]
\centering
\includegraphics[width=6cm]{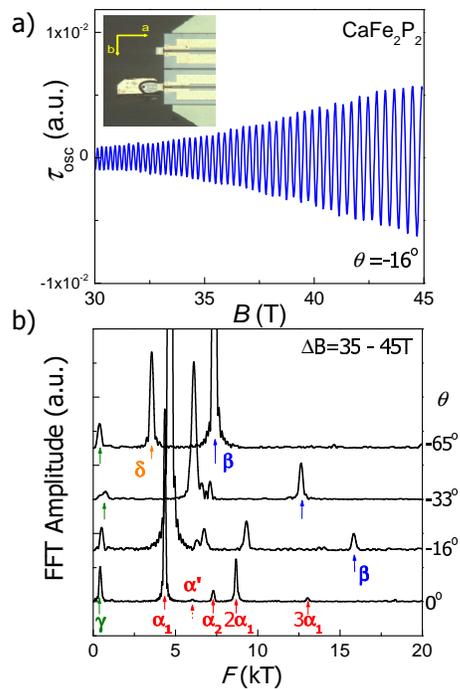}
\caption{(color online) a) Oscillatory part of torque  at
$T=0.4$\,K for a single crystal of CaFe$_2$P$_2$ (sample A).
The inset shows sample A attached to a piezocantilever. b) The Fourier
transform spectra showing the evolution of the extremal areas of the
Fermi surface with the field angle, $\theta$. The positions of different pockets,
$\alpha_1$, $\alpha_2$, $\beta$, $\gamma$, $\delta$ and harmonics
($2\alpha_1$, $3\alpha_1$) as well as a weak peak, $\alpha$' are
indicated by arrows.}\label{fig:rawdatafft}\end{figure}

Here we report a de Haas-van Alphen effect study of the Fermi surface of CaFe$_2$P$_2$ which
is a very close structural analogue
of the cT phase of CaFe$_2$As$_2$.
The isoelectric substitution of As with P does not change the
number of Fe 3d electrons, but enhances P-P hybridization causing the lattice to contract
in the {\it c} direction. Consequently the interlayer P-P distance approaches the
 molecular bond length \cite{GustenauHN97}, just as the As-As distance does in the
cT phase \cite{Yildirim2008a}.
We show experimentally that the Fermi surface of CaFe$_2$P$_2$
is different from the multiband electron-hole structure found in SrFe$_2$P$_2$ \cite{Analytis09}
and other pcnitides, and is formed of
a single warped cylinder and a large three dimensional hole sheet.
We find a isotropic mass enhancements on the electron and hole sheets ($\sim 1.5$).

High quality single crystals of CaFe$_2$P$_2$\, were grown from a Sn flux similar to previous reports
\cite{Analytis09}. The residual resistivity ratio $\rho$(300\,K)/$\rho$(1.8\,K) was measured to be greater than 120.
Torque magnetometry was performed using piezoresistive microcantilevers in high fields 18~T in Bristol (sample B)
and 45~T (sample A) at the NHFML, Tallahassee.  We compare our data with the predictions of bandstructure calculations which we performed using an
augmented plane wave plus local orbital method as implemented in the WIEN2K code \cite{Wien2k}. The experimentally
determined lattice parameters and internal positions  were used for these calculations \cite{StructureDetails}.

Fig.~\ref{fig:rawdatafft}(a) shows the oscillatory part of raw torque signal for a CaFe$_2$P$_2$ crystal up to 45\,T.
The fast Fourier transform (FFT) for several other orientations as the field is rotated from $B||c$
($\theta=0$) towards $B||a$ ($\theta=90^{\circ}$) is shown in Fig.1b.  The FFT frequencies $F$
of oscillatory torque data (in the 1/$B$ domain) are
related to extremal Fermi surface areas by $F=(\hbar/2\pi e) A_k$. Close to $B||c$
the strongest amplitude peak ($\alpha_1$) is at 4.35\,kT and we also observe
several harmonics of $\alpha_1$ (see Fig.1b); we can distinguish a
frequency $\alpha_2 \sim7.3$~kT
and a tiny feature $\alpha$' at 6~kT \cite{mosaic}.
A small pocket $\gamma$ is observed at $ \sim 420$~T;
at higher angles ($\theta=15^{\circ}$) the amplitude of another frequency, $\beta$,
becomes significant and the position of this peak varies strongly with
increasing angle. When we rotate close to $B|| a$
a strong amplitude signal, $\delta$,  corresponding
to an extremal area of 3.2~kT
is found.

\begin{figure}[tbp]
\centering
\includegraphics[width=6cm]{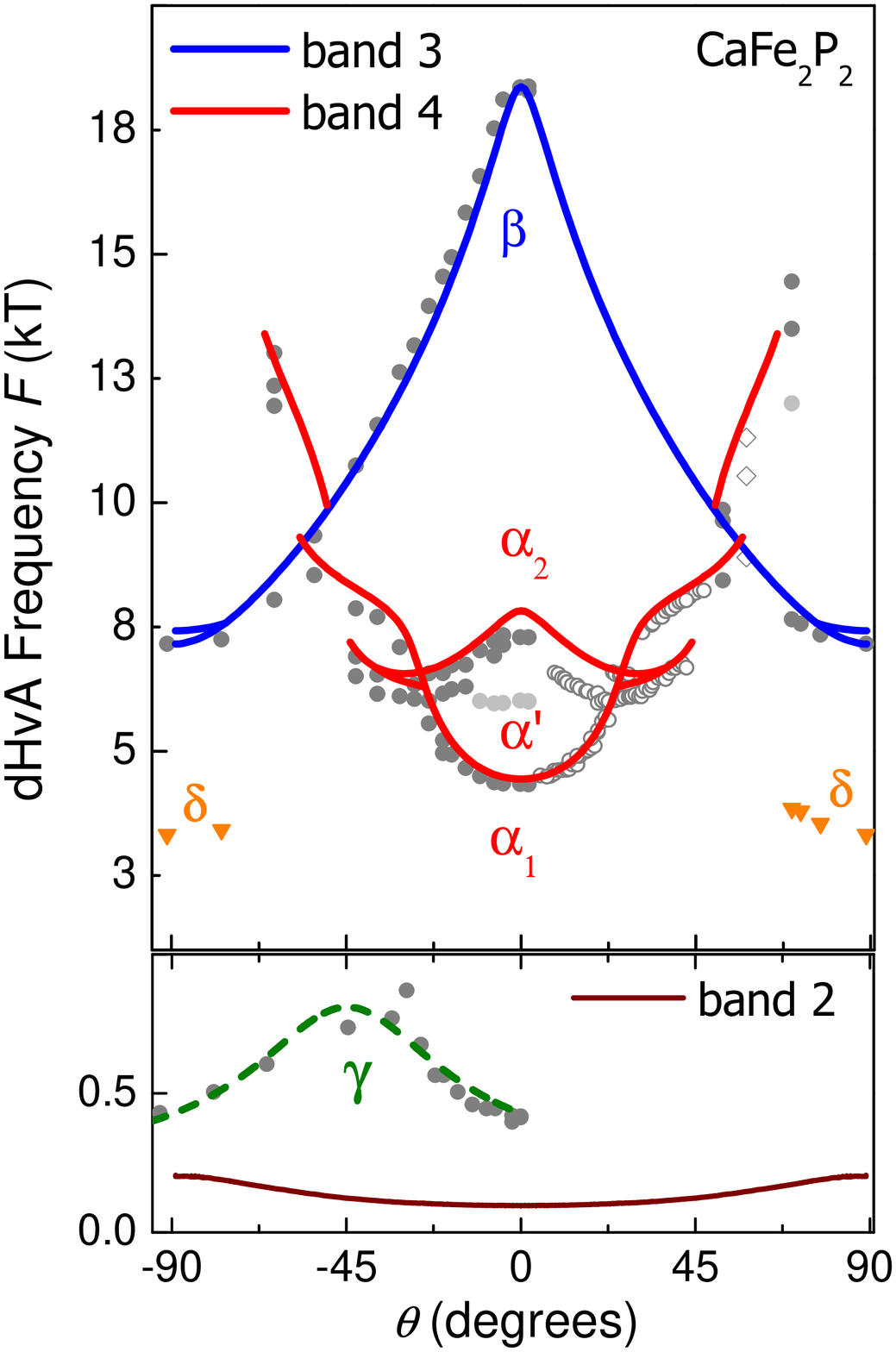}
\includegraphics[width=6.5cm]{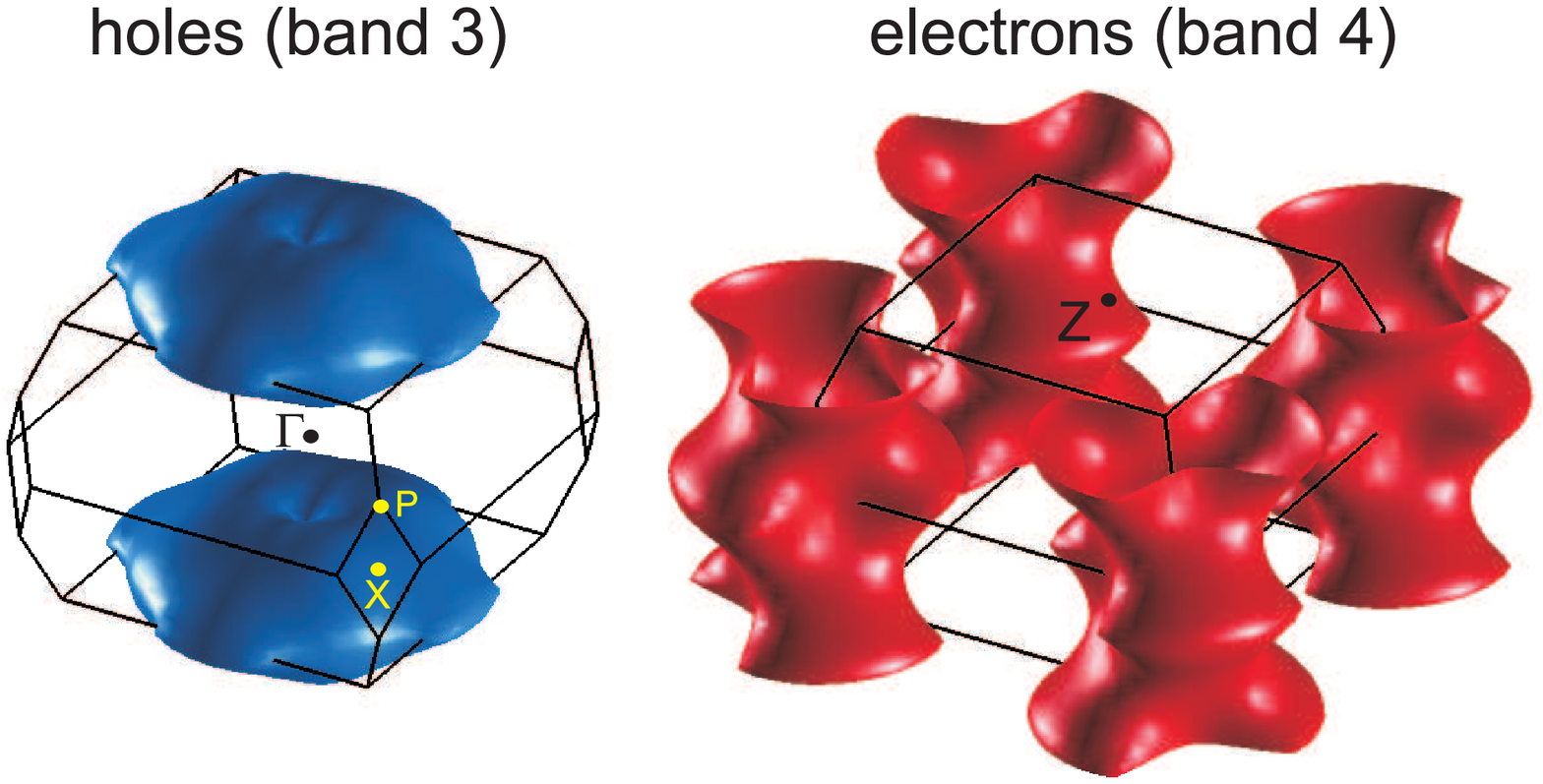}
\caption{(color online) Angle dependence of all observed frequencies
compared with the band structure predictions (solid lines).
Different symbols correspond to sample A (filled
circles) and sample B (open circles).
Solid grey points indicate the position of very weak features.
The bottom panel shows an expanded low frequency scale for the $\gamma$ pocket and
the dotted line is a guide to the eye.
The calculated Fermi surface of CaFe$_2$P$_2$ is also shown.
The solid lines delimits the Brillouin zone
and the Fermi surface sheets
are represented in an extended zone.} \label{fig_rotation_plot}\end{figure}

The angular dependence of the observed dHvA frequencies
allows us to identify the shape of the Fermi surface
sheets from which they originate (see Fig.\
\ref{fig_rotation_plot}). Rotating away from $B||c$
the $\alpha_1$ and $\alpha_2$ orbits display a much stronger angular variation
compared to extremal orbits on a simple cylinder with a weak $k_z$ dispersion
($F$ ($\theta$) $ \sim 1/\cos \theta$),
\cite{Coldea08}.
This suggests that these orbits originate from
a strongly warped cylindrical Fermi surface \cite{mosaic}.
The size of the $\beta$ orbit changes dramatically with $\theta$ suggesting that this Fermi
surface sheet has an prolate ellipsoidal shape with a maximum at
$\theta=0^{\circ}$.
The $\alpha$ orbits are well reproduced in both samples,
but the $\beta$ orbit was only observed in sample A in much higher fields (up to 45\,T).
This can be explained as the impurity damping of the dHvA signal
is proportional to $\exp(-k_F/(\ell B))$
so that higher field are needed to see the
larger $\beta$ orbit ($k_F \propto \sqrt{F}$) for the same mean-free-path
($\ell$).

\begin{figure}
\includegraphics[width=4cm]{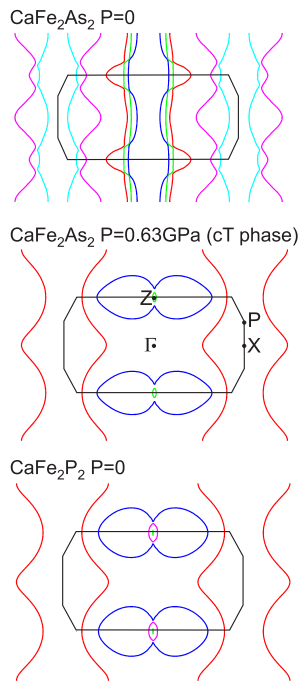}
\caption{(color online) Comparison of the calculated Fermi surface
topology of  CaFe$_2$As$_2$ (tetragonal phase),
 CaFe$_2$As$_2$ (cT phase) and CaFe$_2$P$_2$.
  Slices through the center of the Brillouin zone (solid lines) in the (110) plane are shown.}
 \label{slices}
\end{figure}

We now compared the experimental data in CaFe$_2$P$_2$
to the predictions of the band structure calculations.
The Fermi surface (see Fig.\ \ref{fig_rotation_plot} and Fig.~\ref{slices})
is quasi-three dimensional and is composed of a large
hole sheet in the form of a flat pillow at the top of the zone whereas the electron sheets are strongly
distorted quasi-two-dimensional tubes centered on the zone corners.
There are also two tiny hole pockets centered at Z
(see the bottom panel of Fig.\ref{slices}) containing only $\sim 0.008$ holes compared with
the large hole and electron sheets which each contains $\sim 0.41$ holes/electrons;
if the P position is optimized in the calculation (by minimising the total
energy, $z_P$=0.3890) they disappear.

Fig.\ \ref{fig_rotation_plot}
shows good agreement between data and calculation
in the case of the $\beta$ frequency which corresponds to cross sections on the hole sheet
(band 3) and the $\alpha_1$ and $\alpha_2$ frequencies which correspond to the
minimum and maximal extremal areas of the strongly warped electron cylinder (band 4)
(the degree of warping is roughly $\Delta F/F \sim  50\%$ in CaFe$_2$P$_2$ as
compared with $\sim 23\%$ in SrFe$_2$P$_2$ \cite{Analytis09}).
The complex in-plane and interplane corrugations of
the Fermi surface could give rise to additional branches
not predicted by our calculations.
This could explain the origin of the $\delta$ branch,
which is observed within $\sim 25^\circ$ of $\theta=90^\circ$ ($B || a$)
 (Fig.\ \ref{fig_rotation_plot}) and may account for the small
shift found for $\alpha_2$ as well as the presence of the weak feature at $\alpha$'.
It is worth emphasizing that the
agreement between data and calculations ($\pm 10$meV) for the main sheets is good
in contrast to our findings for LaFePO and  SrFe$_2$P$_2$ \cite{Coldea08,Analytis09} where rigid band
shift of up to 100\,meV were needed to bring the bandstructure into agreement with experiment.

Any small band shifts would mainly affect the
small pockets of the Fermi surface centered at the Z point, as stated earlier.
For example, by shifting the hole bands down the tiny pockets at the Z point disappear
and the large hole pillow transforms into a large torroid \cite{Rourke2008};
when $B || a$ we could expected orbits from extremal areas
on such torroid which may give rise to the $\delta$ branch (which is about half
the size of the $\beta$ orbit close to $B || a$)
(see bottom panel of Fig.~\ref{slices}).
The $\gamma$ frequency
increases to about $\theta=42(3)^{\circ}$ and then
decrease suggesting that it has a short closed cylinder shape (see Fig.\ \ref{fig_rotation_plot}).
By shifting the hole bands up (by $\sim 40$~meV) the small 3D
pocket centered at the Z point (band 2) could be assigned to the
$\gamma$ branch (Fig.\ref{slices}).
Alternatively, $\gamma$ which has a low effective mass ($m^*=0.4 m_e$)
could originate from an Sn impurity phase which has orbits with similar frequencies and masses \cite{Craven69},
but x-ray diffraction measurements does not identify any Sn impurities at the level of $\sim 1\%$.
In any case this pocket, $\gamma$, accounts for only a tiny fraction of holes ($\sim 0.03$)
and we believe it is unlikely to play any major role.

\begin{table}
\caption{Experimental and calculated Fermi surface parameters of CaFe$_2$P$_2$
close to $\theta=0^\circ$ ($B \| c$)
similar to those predicted for CaFe$_2$As$_2$ in the cT phase \cite{Tompsett09}.}
\begin{tabular}{llllllll}
\hline \hline
\multicolumn{4}{c}{Experiment}&\multicolumn{4}{c}{Calculations}\\
 & $F$(kT) &$\frac{m^*}{m_e}$&$\ell$(nm)& Orbit & $F$(kT) &$\frac{m_b}{m_e}$ &$\frac{m^*}{m_b}$\\
\hline
$\alpha_1 (e)$  &4.347    & 2.05(4)        &190&$4_{\rm min}$          &4.439   &1.38     & 1.49(3)\\
$\alpha_2 (e)$  &7.295    & 3.48(7)        &71&$4_{\rm max}$           &7.837   &2.40     & 1.45(3)\\
$\beta    (h)$  &18.360   & 4.0(2)        &86&$3_{\rm min}$            &18.407  &2.65     & 1.51(8)\\
\hline
\hline
\end{tabular}
\label{Tablemassfreq}
\end{table}

Fig.\ \ref{slices} shows a comparison between the Fermi surface of CaFe$_2$As$_2$ in the tetragonal phase ($c/a$=2.98),
CaFe$_2$As$_2$ in the cT phase ($c/a$=2.66 at $P$=0.63~GPa) and  CaFe$_2$P$_2$ ($c/a$=2.59).
As mentioned before, it is clear that there is a remarkable similarity between
CaFe$_2$P$_2$ and the cT phase of
CaFe$_2$As$_2$.
Thus in CaFe$_2$As$_2$ applying {\em chemical pressure}
(by the isoelectronic substitution of As by P)
is equivalent to {\em applied hydrostatic pressure}, as found in
other ternary pnictides \cite{Kimber2009}.
This state of reduced  $c/a$ ratio
has a different Fermi surface topology compared to CaFe$_2$As$_2$ or SrFe$_2$P$_2$ ($c/a$=3.03).
Yildirim \cite{Yildirim2008a} has argued that the cT phase of CaFe$_2$As$_2$ occurs
when, by reducing the Fe moment, the Fe-As bonding weakens and
the (inter and intra-planar) As-As bonding gets stronger causing the significant strong
reduction in the $c$ axis \cite{KreyssigGLSZLBTNNLPAHMCG08}.
Similarly, in non-magnetic phosphides,
the reduction in the $c$-axis (or the $c/a$ ratio)
results in an increase P-P hybridization between
pnictogen ions along the $c$ direction (close to the single bond distance) \cite{GustenauHN97}.
The spacer between the iron layers (Sr or Ba) limits
the degree of this hybridization between layers and such
a state with {\em strong pnictogen bonding} is unlikely to occur
\cite{Zhang_BS08,Kimber2009}.

%

The effective masses, $m^*$, of CaFe$_2$P$_2$ for
each Fermi surface orbit extracted from the temperature dependence of the
dHvA signals, using the conventional Lifshitz-Kosevich formula \cite{Shoenberg},
are shown in Table \ref{Tablemassfreq}
and compared to the corresponding bandstructure values.
The masses for the electron and hole sheets are enhanced
by the same factor of $\sim 1.5$ in contrast to the sheet dependent
variation of the enhancement observed recently in SrFe$_2$P$_2$
(which has two electron and two hole Fermi surface pockets).
By comparing the
quasiparticle enhancement on the electron sheets (which have the
largest mean free path and often match better to the bandstructure
calculations being less sensitive to structural changes compared with the hole pockets)
we observe that in LaFePO the
average enhancement is 2.38 (2.2 for inner and 2.54 for outer
pocket) \cite{Coldea08} whereas in SrFe$_2$P$_2$ is 1.85 (1.6 for
inner and 2.1 for the outer electron sheet)\cite{Analytis09}.
The conventional electron-phonon coupling, $\lambda_{e-ph}$,
in the cT phase of CaFe$_2$As$_2$ is calculated to be 0.23 \cite{Yildirim2008a} but such
calculations for CaFe$_2$P$_2$ are not yet available. Due to their structural and
electronic similarities it is likely that in CaFe$_2$P$_2$ a mass
enhancement of 1.23 could be due to a conventional electron-phonon coupling, but other
effects could also be important \cite{lambda}.
The further anisotropic
mass enhancement observed in LaFePO and possibly in SrFe$_2$P$_2$ could have another
origin and may be related to the nesting of the Fermi surface.
The mean free paths, $\ell$, of the
minimum electronic orbit is a factor 2 larger than that of large hole sheet and of the maximum electronic orbit. This
suggests both anisotropy in scattering between electron and holes but also along $k_z$, as also
observed in SrFe$_2$P$_2$ and LaFePO \cite{Analytis09,Coldea08}.

In the case of the superconducting LaFePO and SrFe$_2$P$_2$ \cite{Coldea08,Analytis09}
the energies in the band structure were shifted in opposite direction for
the electron and hole pockets to match up the experimental data.
These asymmetric band shifts are suggested to result from geometric nesting
\cite{Ortenzi09}.
In the present measurements on CaFe$_2$P$_2$ no band shifts were required  ($\sim \pm 10$ meV) to
achieve agreement with experiment and
considering this model it would imply the absence of nesting in CaFe$_2$P$_2$
(or the lack of long-range order in the cT phase of CaFe$_2$As$_2$ \cite{GoldmanKPPALNKCLSLPBNCHM09}.)

In conclusion, we have experimentally determined the Fermi surface of CaFe$_2$P$_2$ which is closely related to the
collapsed tetragonal phase of CaFe$_2$As$_2$.
We find that the Fermi surface is composed of a single
highly dispersive electron cylinder at the zone corners and a
large three dimensional pillow shaped hole surface. The
mass enhancement due to many-body interactions is
isotropic ($\sim 1.5$)
and may be dominated by the electron-phonon coupling.
The features of this Fermi surface which does not fulfill
nesting condition is likely to be shared by the non-magnetic cT phase
of CaFe$_2$As$_2$ and other ternary pcnitides
with reduced $c/a$ ratio and may explain
why superconductivity is absent in such a state.

We thank M. Haddow and E. Yelland for technical support.
We acknowledge financial support from the Royal Society, EPSRC, and EU 6th Framework contract
RII3-CT-2004-506239. Work at Stanford was supported by the U.S. DOE, Office of Basic Energy Sciences under contract
DE-AC02-76SF00515. Work performed at the NHMFL in Tallahassee, Florida, was supported by NSF Cooperative Agreement No.
DMR-0654118, by the State of Florida, and by the DOE.


\begin{thebibliography}{24}
\bibitem[{mos(????)}]{mosaic}
\bibinfo{note}{The X-ray diffraction shows that sample A contain a single
  domain with a ($\sim 1^\circ$) mosaic spread and about one percent of
  another phase missaligned in the ($ac$) plane}.
\bibitem[{lam(????)}]{lambda}
\bibinfo{note}{Many-body interactions could be responsible
for a futher enhancement of the electron-phonon
  interaction due the strong polarization of pictogen
  ions \cite{Kulic09}.}
\end{thebibliography}

\end{document}